\documentclass[10pt,a4paper]{article}
\newcommand\BK{Benjamin Kowarsch}

\usepackage[top=2.4cm, bottom=2.4cm]{geometry} 
\usepackage{enumitem} 
\usepackage{listings} 
\usepackage{xcolor} 
\usepackage{url} 

\usepackage{courier}
\usepackage{lmodern}
\makeatletter
\newcommand{\verbatimfont}[1]{\def\verbatim@font{#1}}
\makeatother

\usepackage{fancyhdr} 
\renewcommand{\emph}[1]{\textbf{\textit{#1}}}

\lstdefinestyle{modula2}{
  language=Modula-2,
  frame=none,
  basicstyle=\fontfamily{pcr}\selectfont,
  keywordstyle=\textbf,
  commentstyle=\italicgray
}

\newcommand\boldgray[1]{\textcolor{gray}{\textbf{#1}}}
\newcommand\italicgray[1]{\textcolor{gray}{\textit{#1}}}
\newcommand\middledots{\textperiodcentered\textperiodcentered\textperiodcentered}

\usepackage[automake, nonumberlist, style=altlist]{glossaries}
\makeglossaries
\newglossaryentry{offending facility}{
  name=offending facility, plural=offending facilities,
  description={A language facility that is (a) outdated, harmful or bad habit
  forming in general,\linebreak or (b)~violates any of the design principles
  in section \ref{design-principles}} in particular
}
\newglossaryentry{API}{
  name=API,
  description={Application programming interface}
}
\newglossaryentry{foreign definition module}{
  name=foreign definition module,
  description={A definition module that specifies an interface to a
  \gls{foreign API}}
}
\newglossaryentry{compiler directive}{
  name=compiler directive,
  description={A directive within the source to instruct a language processor
  how to process the input}
}
\newglossaryentry{ETH}{
  name={ETH},
  description={Eidgen\"{o}ssisch Technische Hochschule
  -- Swiss Federal Institute of Technology}
}
\newglossaryentry{foreign identifier}{
  name=foreign identifier,
  description={An identifier of a foreign API}
}
\newglossaryentry{foreign API}{
  name=foreign API,
  description={An API implemented in a language other than Modula-2}
}
\newglossaryentry{non-semantic compiler directive}{
  name=non-semantic compiler directive,
  description={A compiler directive that does not alter the semantics of the
  source code}
}
\newglossaryentry{semantic compiler directive}{
  name=semantic compiler directive,
  description={A compiler directive that alters the semantics of the
  source code}
}

\usepackage{pifont}

\title{On the Maintenance of Classic Modula-2 Compilers}
\author{\BK, Modula-2 Software Foundation}
\date{\small{September 2018 (arXiv.org preprint)}}
\makeatletter
\let\Title\@title
\makeatother
\makeatletter
\let\Author\@author
\makeatother

\begin{document}
\verbatimfont{\small\fontfamily{lmtt}\selectfont}
\maketitle

\begin{abstract}
The classic Modula-2 language was specified in \cite{Wirth78} by N.Wirth at
\Gls{ETH} Z\"{u}rich in 1978. The last revision \cite{Wirth88} was published in
1988. Many computer science books of that era used Modula-2 in programming
examples. Many of these are still valuable resources in computer science
education today. To compile and run the examples therein, it is essential to
have compilers available that follow the classic Modula-2 language definition
and run on modern computer hardware and operating systems. Although most
Modula-2 compilers of that era have disappeared, a few have since been
re-released under open source licenses. Whilst the original authors have long
ceased work on these compilers, new maintainers have stepped in. This paper
gives recommendations for maintenance on classic Modula-2 compilers while
balancing the aim to modernise with the need to maintain the capability to
compile programming examples in the literature with minimal effort. 
Nevertheless, the principles, methods and conclusions presented are adaptable
to maintenance on other languages.
\end{abstract}

\section{Methodology}

\subsection{Design Principles}
\label{design-principles}

The following design principles strongly influenced the recommendations in
this paper:

\subsubsection{Single Syntax Principle (SSP)}
\label{SSP}

There should be one and \emph{only one} syntax form to express any given
concept \cite{Dijkstra78}.

\subsubsection{Literate Syntax Principle (LSP)}
\label{LSP}

Syntax should be chosen for readability and comprehensibility by a human
reader \cite{Knuth84}.

\subsubsection{Consistency of Syntax Principle (COSP)}
\label{COSP}

Syntax should be consistent. Analogous concepts should be expressed by
analogous syntax.

\subsubsection{Principle of Least Astonishment (POLA)}
\label{POLA}

Of any number of possible syntax forms or semantics, the one likely to cause
the least astonishment for a human reader should be chosen and the alternatives
should be discarded \cite{Geoffrey87}.

\subsubsection{Single Responsibility Principle (SRP)}
\label{SRP}

Units of decomposition, such as modules, classes, procedures and functions
should have a single focus and purpose \cite{Martin09}.

\subsubsection{Principle of Information Hiding (POIH)}
\label{POIH}

Implementation specific details should always be hidden, public access should
be denied \cite{Parnas72}.

\subsubsection{Safety Perimeter Principle (SPP)}
\label{SPP}

Facilities that undermine the safety otherwise safeguarded within the language
should be segregated from other facilities \cite[ch.29]{Wirth88}. Their
use should require an explicit expression of intent by the author and be
syntactically recognisable so as to alert the author, maintainer and reader
of the possible implications. This applies and extends the principle of least
privilege \cite{Saltzer74}.

\subsection{Maintenance Objectives}

The primary objectives for the recommendations in this paper are:

\renewcommand{\labelenumi}{(\arabic{enumi})}
\begin{enumerate}[leftmargin=!, labelindent=-0.75em, itemindent=0em]
\item to remove facilities that are harmful, outdated or violate
any of [\ref{design-principles}]
\item to resolve ambiguities in \cite{Wirth88} and incompatibilities between
implementations
\item to maintain the capability to compile programming examples in the
literature with \raggedright{minimal~effort}
\end{enumerate}

\subsubsection{Weighing Objectives by Impact}

\noindent The objectives given above may from case to case conflict with one
another. Which objective should be given preference in the event of a conflict
depends on the following factors:

\begin{enumerate}[itemindent=-0.75em]
\item the severity of the \gls{offending facility}
\item the estimated frequency of use of the \gls{offending facility}
\item the effort to update sources impacted by change or removal of the
\gls{offending facility}
\end{enumerate}

\noindent The greater the severity of an \gls{offending facility}, the stronger
is the case for \emph{change} or \emph{removal}; the lower the estimated
frequency of use and the less the effort to update impacted sources, the
stronger the case for \emph{change} or \emph{removal}. In the event that
the estimated frequency of use is high and the effort to update impacted
sources is significant, \emph{deprecation} is recommended.

\subsection{Mitigation Methods}

Terms of mitigation methods used in this paper have well defined meanings:

\subsubsection{Warning}

A warning should be issued for each and every use of the
\gls{offending facility}.

\subsubsection{Change}

The \gls{offending facility} should be replaced with a proposed alternative.

\subsubsection{Deprecation}

A compiler switch to enable and disable the \gls{offending facility} should
be provided and it should be disabled by default. When it is enabled, a
deprecation warning should be isued for each and every use of the
\gls{offending facility}.

\subsubsection{Removal}

Support for the \gls{offending facility} should be removed altogether.\\

\par\noindent From an educational perspective, the availability of
\glspl{offending facility} promotes bad habits. Replacement or removal
is therefore generally preferable to warning or deprecation.

\subsubsection{Transformation}

To be used in combination with any of change, deprecation or removal.
A conversion program should be provided that transforms source code that uses
\glspl{offending facility} into semantically equivalent source code that
complies with the recommendations given in this paper.

\section{Lexis}

\subsection{Octal Literals}

Support for octal literals should be \emph{removed}.

\subsubsection{Rationale}

The use of octal numbers has long been outdated. Further, the \verb|B| and
\verb|C| suffixes used to denote octal literals are also legal digits within
hexadecimal literals, which is confusing to human readers of the source code as
it violates POLA [\ref{POLA}] and it unnecessarily complicates lexing.

\subsubsection{Substitution}

The built-in \verb|CHR()| function can be used instead without penalty as it is
evaluated at compile time for constant arguments. The function accepts decimal
and hexadecimal arguments. Code that uses the \verb|CHR()| function instead of
octal literals can always be compiled on any classic Modula-2 compiler,
regardless of whether octal literals are recognised or not.

\subsubsection{Backwards Compatibility}

Preferably, \emph{transformation} should be used to replace all occurences of
octal literals in existing Modula-2 sources. Alternatively, octal literals
could be \emph{deprecated} instead.

\subsection{Synonym Symbols}

Support for synonym symbols \verb|<>|, \verb|&| and \url{~} should be
\emph{removed}.

\subsubsection{Rationale}

The availability of alternative symbols violates SSP (\ref{SSP}) and they are
are notably \emph{absent} from the grammar in \cite[pp.156-157]{Wirth88}.

The inequality operator symbol \verb|#| is preferable to its synonym \verb|<>|
because it resembles the mathematical inequality symbol \footnotesize
\raisebox{0.35ex} {$\neq$} \normalsize and as a single character symbol it 
simplifies lexing. Reserved words \verb|AND| and \verb|NOT| are preferable to
their respective synonyms \verb|&| and \url{~} because of consistency: While
there are synonyms for \verb|AND| and \verb|NOT|, there is none for \verb|OR|.
Although \verb!|! could have been used to denote \verb|OR|, it is already used
to separate case labels and would have caused ambiguity.

\subsubsection{Substitution}

Symbol \verb|#| can be used in place of \verb|<>|, while \verb|AND| can be used
in place of \verb|&| and \verb|NOT| in place of \url{~}.

\subsubsection{Backwards Compatibility}

Preferably, \emph{transformation} should be used to replace all occurences of
synonym symbols in existing Modula-2 sources. Alternatively, synonym symbols
could be \emph{deprecated} instead.

\subsection{Non-Semantic Compiler Directives}

Whilst \glspl{compiler directive} are implementation defined, the delimiters of
\glspl{non-semantic compiler directive} should not be. They should be denoted
by an opening \verb|(*$| and a closing \verb|*)| delimiter.

\subsubsection{Rationale}

Although \cite[p.18]{Wirth88} mentions \verb|(*| and \verb|*)| as delimiters
for both comments and directives, it does not specify their use in any
normative manner. Some compiler implementors have taken this to mean that
the delimiters are implementation defined. Several compilers still in use
today do not follow the convention, for example the venerable MOCKA compiler
\cite{MOCKA}.

As a result, source code with \glspl{compiler directive} is often not portable
across different implementations. This is unfortunate, because non-semantic
directives can safely be ignored in the event that they are not supported.
Even where an implementation issues warnings about unrecognised directives,
to be able to identify a directive as such in the first place, a common
delimiter convention needs to be followed across implementations.

\subsubsection{Backwards Compatibility}

Non-compliant directives within existing source code should be replaceable with
minimal effort using regular expressions and a filter program such as
\verb|sed| or \verb|awk|.

\subsection{Semantic Compiler Directives}

\Glspl{semantic compiler directive} should \emph{not} be denoted by the same
delimiters used for \glspl{non-semantic compiler directive}.

\subsubsection{Rationale}

Whilst \glspl{non-semantic compiler directive} can safely be ignored,
\glspl{semantic compiler directive} cannot. A \gls{semantic compiler directive}
that is not supported by an implementation must be reported as an error.
Consequently, an implementation needs to be able to distinguish between
semantic and non-semantic directives. In order to do so, different delimiters
must be used.

\subsubsection{Substitution}

Some Modula-2 compilers have used a \verb|%|
prefix to denote \glspl{compiler directive}, in particular for conditional
compilation, for example the  MOCKA compiler \cite{MOCKA}. We recommend to
follow that convention to denote \glspl{non-semantic compiler directive}
if any are provided. The \verb|%|
symbol is not otherwise used in the language.

\subsubsection{Backwards Compatibility}

Non-compliant directives within existing source code should be replaceable with
minimal effort using regular expressions and a filter program such as
\verb|sed| or \verb|awk|.

\section{Syntax}

\subsection{Multi-Dimensional Arrays}

Inconsistent definition or declaration of multi-dimensional arrays should
trigger \emph{warnings}.

\subsubsection{Rationale}

\cite[p.138]{Wirth88} specifies alternative syntax forms for multi-dimensional
array type declaration:

\lstset{style=modula2}
\begin{lstlisting}
TYPE Matrix = ARRAY [0 .. Cols], [0 .. Rows] OF REAL;
\end{lstlisting}

\noindent is an abbreviation of and thus equivalent to 
\lstset{style=modula2}
\begin{lstlisting}
TYPE Matrix = ARRAY [0 .. Cols] OF ARRAY [0 .. Rows] OF REAL;
\end{lstlisting}

\par\noindent The availability of alternative syntax forms violates
SSP [\ref{SSP}] and COSP [\ref{COSP}]. The more dimensions an array type has,
the more preferable the abbreviated form becomes. However, it is more effort
to \emph{remove} or \emph{deprecate} the long form. The impact of
\emph{removal} or \emph{deprecation} on existing sources with
multi-dimensional arrays would likely be high. It is less effort and
sufficient to modify an implementation to detect and warn about mixed use
in any given compilation unit.

\subsubsection{Backwards Compatibility}

The proposed mitigation does not impact the compatibility of legacy sources.

\subsection{Local Modules}

Local Modules should be \emph{deprecated}.

\subsubsection{Rationale}
If there is sufficient reason to delegate certain responsibilities of a library
module to a local module, then there is also sufficient reason to delegate
those responsibilities to a separate library module. There is no reason why a
local module should be chosen over a separate library.

Furthermore, a local module within a program or library module unnecessarily
increases the line count of the module and thus reduces its readabililty and
maintainability. It runs counter to the very rationale of decomposing source
code into separate modules in the first place.

Finally, a test environment for testing a local module must necessarily be
provided within the hosting module.  This further increases clutter and poses
the question whether to release the module with or without the embedded test
environment. By contrast, a proper library module can be imported into any
number of lexically independent test environments.

\subsubsection{Substitution}
Local modules should be removed from their enclosing module and provided as
proper library modules with their own definition and implementation parts.

\subsubsection{Backwards Compatibility}

Backwards compatibility can be provided via compiler switch.

\subsubsection{Private-Use Aspect}

The private-use aspect of local modules is lost when they are removed from
their host modules and converted into library modules. However, this aspect
is useful when the local module's API is considered unstable and subject to
frequent changes. A satisfactory solution can be implemented easily by
introducing a \gls{non-semantic compiler directive} that specifies a
library's intended client modules and causes the compiler to issue warnings
whenever such a library is imported by a module other than its designated
client modules. An example is given below:

\lstset{style=modula2}
\begin{lstlisting}[escapechar=~]
DEFINITION MODULE PrivateLib; ~\boldgray{(*\$CLIENTS=FooLib, BarLib, BazLib*)}~
(* ATTENTION !!! This module is intended for private use only.
 * Its API is subject to frequent change without prior notice.
 * The compiler will issue a warning when it is imported from
 * any other than the designated client modules. *)
 ~\middledots~
END PrivateLib.
\end{lstlisting}

\subsection{Unary Minus}

Unary minus should only be permitted before a factor.

\begin{verbatim}
simpleExpression :=
  ( '+' )? term ( addOp term )* | '-' factor
  ;
\end{verbatim}

\noindent instead of

\begin{verbatim}
simpleExpression :=
  ( '+' | '-' )? term ( addOp term )*
  ;
\end{verbatim}

\subsubsection{Rationale}

\cite{Wirth88} does not state the scope of the unary minus operator other than
through the grammar.\\

\noindent An expression of the form
\begin{lstlisting}
- a * b + c
\end{lstlisting}

\noindent could be interpreted \emph{mathematically} correct
\begin{lstlisting}
(-a) * b + c
\end{lstlisting}

\noindent or \emph{grammatically} correct
\begin{lstlisting}
-(a * b + c)
\end{lstlisting}

The former interpretation conforms to mathematical convention. The latter can
be deduced from the grammar in \cite[pp.156-157]{Wirth88} and is intended to
be the correct interpretation\footnote{The author obtained clarification on
the precedence of unary minus from Prof. Wirth by email.}.

\newpage

\noindent However, this violates POLA [\ref{POLA}] and some implementations
therefore follow the mathematically correct interpretation, for example the
ACK compiler \cite{ACK} and the MOCKA compiler \cite{MOCKA}. Requiring a
factor after a unary minus forces the use of parentheses whenever there is
more than one factor to follow which makes programmer intent explicit.

\subsubsection{Backwards Compatibility}

Unary minus is an infrequently used operation. When it is used, it is usually
in a form that is compliant with the syntax proposed in this paper. In the
unlikely event that it is not in a compliant form, the source code is ambiguous
and needed to be fixed anyway. It may well have been written for and tested
with a compiler that uses a different interpretation than the compiler now
used to compile the sources. The proposed mitigation will thereby help
identify any non-compliant occurrences which can then be corrected with
minimal effort.

\section{Pervasives}

\subsection{Conversion}

Pervasive function \verb|VAL()| should be \emph{changed}
to cover all numeric conversions and numeric conversions only.
Pervasive functions \verb|FLOAT()| and \verb|TRUNC()| should then be
\emph{deprecated}. Any additional non-standard conversion functions such as
\verb|INT()|, \verb|CARD()| and \verb|LFLOAT()| should be \emph{removed}.

\subsubsection{Rationale}

\cite[p.150]{Wirth88} specifies pevasive conversion function \verb|VAL()|
which covers all possible use cases for safe type conversion between whole
number types. There is no reason why \verb|VAL()| could not also cover all
possible use cases for conversions between whole and real number
types. By contrast, \texttt{CHR()} and \texttt{ORD()} are not
conversion functions\footnote{\texttt{CHR()} performs a lookup while
\texttt{ORD()} returns a property.} and they should not be duplicated by
\verb|VAL()|. Duplication violates SSP [\ref{SSP}].
 
Furthermore, function \verb|FLOAT()| is confusingly named since the type it
converts to is \verb|REAL| and there is no type \verb|FLOAT|. \cite{Wirth88}
does not specify how real number types are to be implemented. They need not be
implemented as floating point numbers but could be implemented as binary coded
decimals, for instance. The naming of \verb|FLOAT()| is thus
misleading and violates COSP [\ref{COSP}] and POLA [\ref{POLA}].

Finally, function \verb|TRUNC()| violates SRP [\ref{SRP}] as it represents
both a conversion function and the mathematical function $trunc(x)$.
Its name is misleading as it does not suggest conversion, violating
POLA [\ref{POLA}]. Conversion should be the responsibility of function
\verb|VAL()| and truncation the responsibility of a math function
\verb|trunc()| to be provided in a library such as \verb|MathLib| and
return the same type as its argument, it should not perform any type
conversion.

\subsubsection{Substitution}

Once function \verb|VAL()| has been extended to provide type conversion
between whole and real number types, any invocation of functions \verb|FLOAT()| and \verb|TRUNC()| within existing source code may be safely
replaced with an invocation of \verb|VAL()|.

\subsubsection{Backwards Compatibility}

Backwards compatibility can be provided via compiler switch.

\subsubsection{Implementation}
It should be noted that function \verb|VAL()| is not generally implemented as a
single function, nor should it be. Instead, it is usually and preferably
implemented as a built-in macro, resolved at compile time to a function
internal to the compiler that is specific to the source and target types of its
arguments, since the types are known at compile time. \verb|VAL()| thereby
simplifies the user interface without incurring the penalty of increasing the
complexity of its implementation. The specific conversion functions need to be
implemented within the compiler anyway, but they should not be individually
exposed in the user interface.

\section{Semantics}

\subsection{Exported Variables}

Variables defined in definition parts should be exported \emph{read-only}. 

\subsubsection{Rationale}

Allowing a client module to write to imported variables violates POIH
[\ref{POIH}]. And indeed, \cite[p.88]{Wirth88} explicitly states that
``imported variables should be treated as `read-only' objects''.

\subsubsection{Substitution}

A setter procedure can be defined and exported by the same module if needed.

\lstset{style=modula2}
\begin{lstlisting}
DEFINITION MODULE Foo;

VAR bar : Bar; (* read-only *)

PROCEDURE SetBar ( value : Bar );

END Foo.
\end{lstlisting}

\subsubsection{Backwards Compatibility}

Considering that Wirth explicitly stated to treat imported variables as
``read-only" objects, noone should have written any code that treats them
as mutable objects. Unfortunately, there will be code written by careless
programmers who didn't follow his advice. The cautious maintainer may therefore
consider providing a compiler switch to enable and disable write-access to
imported variables, which should then be disabled by default.

\subsection{Pointer Variables}
All pointer variables should be initialised to \verb|NIL| and deallocation
should reset them to \verb|NIL|.

\subsubsection{Rationale}
The two primary paradigms of Modula-2 are (a) program decomposition through
data encapsulation and information hiding, and (b) reliability through type
safety. Opaque pointer types are the primary instrument through which the
former is achieved. The ability to test whether an opaque pointer type has
been allocated or deallocated is central to achieving the latter.

\subsubsection{Backwards Compatibility}

The proposed mitigation does not impact the compatibility of legacy sources.

\subsection{Unsafe Facilities}

Any facility that bypasses the type safety provided by the language should be
disabled by default.

\subsubsection{Unsafe Facilities provided by SYSTEM}
\cite[p.121]{Wirth88} defines pseudo-module \verb|SYSTEM| as a container for
unsafe facilities. The facilities provided therein are only enabled by import.
This is desirable as it sensitises programmers to the fact that the facilities
are unsafe and thereby discourages their use.
 
\subsubsection{Unsafe Facilities NOT provided by SYSTEM}
Unfortunately, not all unsafe facilities are provided through pseudo-module
\verb|SYSTEM|. Two unsafe facilities are provided in the language core and are
enabled by default without import:

\begin{enumerate}[itemindent=-0.75em]
\item Unsafe type transfers \cite[p.119]{Wirth88}, also known as
\emph{type casts}
\item Records with variant parts \cite[p.71, p.138]{Wirth88}, also known as
\emph{variant records}
\end{enumerate}

\par\noindent This is a violation of SPP [\ref{SPP}] and given its implications
for program safety it is unacceptable. In order to mitigate this situation,
support for type casts and variant records should be disabled by default and
require enabling by compiler switch.

\subsubsection{Ideal World Scenario}

In an ideal world, the type cast syntax would be replaced with a \verb|CAST()|
function provided by \verb|SYSTEM| as in ISO Modula-2 \cite{ISO96}, and variant
records with type safe extensible records as in Oberon \cite{Wirth90}. However,
this would constitute a substantial language revision and as such go beyond the
scope of the maintenance aspect of this paper. It would also run counter to the
objective to allow the compilation of programming examples in the literature
with minimal effort.

\section{Language Extensions}

\subsection{Availability}
All language extensions should be disabled by default and only enabled by
compiler switch.

\subsubsection{Rationale}
Disabling language extensions by default aids and promotes writing of portable
source code.

\subsection{Smallest Addressable Unit}

Under no circumstances should any implementation change the definition of type
\verb|SYSTEM.WORD|. However, an implementation targeting an architecture where
the smallest addressable storage unit is eight bits wide should provide an
alias type \verb|BYTE| in module \verb|SYSTEM| as follows:

\lstset{style=modula2}
\begin{lstlisting}
TYPE BYTE = WORD;
\end{lstlisting}

\subsubsection{Rationale}

\cite[p.153]{Wirth88} specifies \verb|SYSTEM.WORD| as the smallest addressable
unit\footnote{When the first report on Modula-2 was written it was still common
to call the smallest addressable storage unit a \emph{word}. Since then the
terminology has changed and it has become common to use the term \emph{byte}
instead.}. Whilst the report permits provision of \emph{additional} facilities
in \verb|SYSTEM|, it does not permit \emph{alterations} of \verb|SYSTEM|
facilities specified in the report.

It would thus be permissible to provide an additional type
\verb|SYSTEM.MACHINEWORD| that represents a machine word larger than the
smallest addressable unit, but it is not permissible to change
\verb|SYSTEM.WORD| to represent a machine word that is not the smallest
addressable unit.

\subsection{Foreign Identifiers}

An implementation that provides a means to interface to \glspl{foreign API},
should also allow the use of \glspl{foreign identifier}. Such an identifier may
contain one or more dollar \verb|$| and/or lowline \verb|_| characters. The use
of \glspl{foreign identifier} for other purposes should be discouraged. 
Support should be disabled by default and enabled by compiler switch when
needed.

To avoid any collission with name mangled symbols generated by the Modula-2
compiler, no consecutive and no trailing dollar signs should be permitted; no
leading, no consecutive and no trailing lowlines should be permitted. Further,
dollar signs and lowlines should not be permitted within module identifiers. 

\subsubsection{Rationale}

Operating system \Glspl{API} and other \glspl{foreign API} often include
variables and procedures with identifiers that include dollar \verb|$|
(predominantly on the VMS operating system) and lowline \verb|_|
(predominantly on Unix systems and C \glspl{API} in general).

\subsection{Foreign Definition Modules}

An implementation that provides a means to interface to \glspl{foreign API},
should use a \gls{non-semantic compiler directive} to mark a definition module
as a \gls{foreign definition module}.

\subsubsection{Rationale}

\cite{Wirth88} does not mention \glspl{foreign definition module}. As a result,
implementors have invented their own syntax to mark a definition module as
foreign, and this varies between implementations.

However, the corresponding implementation could in principle be done in
Modula-2. The marking of a definition module as foreign does not alter its
semantics. Consequently, any such marking constitutes de-facto a
\gls{non-semantic compiler directive}.

\subsubsection{Proposed Syntax}

The directive should be placed after the module header, its proposed syntax
is as follows:

\begin{verbatim}
ffiPragma :=
  '(*$' ffiPragmaKey '=' '"' foreignAPI '"' '*)'
  ;
  
ffiPragmaKey :=
  'F' | /* if implementation uses single-letter keys */
  'FFI' /* if implementation uses multi-letter keys */
  ; 

foreignAPI :=
  'ASM' | 'C' | 'Fortran' | 'Pascal' | ...
  ;
\end{verbatim}

\section{Miscellaneous}

\subsection{Filename Suffixes}

Modula-2 implementations should \emph{only} recognise input files with
suffixes \verb|def| and \verb|mod|.

\subsubsection{Rationale}

Neither \cite{Wirth78} nor \cite{Wirth88} mention filename suffixes for
Modula-2 source files. A de-facto standard has been established by the
compilers distributed by \Gls{ETH} Z\"{u}rich: Suffix \verb|def| is used for
definition module files and suffix \verb|mod| for implementation and program
module files.

The absence of any mentioning of filename suffixes in \cite{Wirth78} and
\cite{Wirth88} has been taken by some implementors as an invitation to
define their own non-standard filename suffixes.

The use of non-standard suffixes leads to unnecessary effort when compiling
sources across different implementations. Moreover, it has led to conflicts
with other notations in and inconsistent support for Modula-2 by various text
editors and source code renderers.

\subsection{Pre-Revision Compilers}
An implementation that follows the original monograph \cite{Wirth78} or the
second edition \cite{Wirth83} should be updated to follow the third
\cite{Wirth85} or, preferably the fourth \cite{Wirth88} edition.

\subsubsection{Rationale}
Prior to \cite{Wirth85} a definition module was required to have an export
list. This was unnecessary duplication since the purpose of a definition
module is to export all its contents.

\newpage

\printglossary[title=Definitions, toctitle=Definitions]

\bibliographystyle{alpha}
\bibliography{Classic-M2-Compiler-Maintenance}
\end{document}